\documentclass[aps,prd,groupaddress,floatfix,tighten,nofootinbib,showpacs,amsfonts,superscriptaddress]{revtex4-2}
\usepackage{url}
\usepackage{xcolor}
\usepackage{hyperref}
\usepackage[latin1]{inputenc}
\usepackage{appendix}
\colorlet{shadecolor}{gray!15}
\definecolor{greenLinks}{rgb}{0, 0.6, 0}
\definecolor{blueLinks}{rgb}{0, 0, 0.6}
\definecolor{redLinks}{rgb}{0.6, 0, 0}
\definecolor{tempText}{rgb}{0.55, 0.10,0.67}
\definecolor{eprintLinks}{rgb}{0.4, 0.4, 0.4}
\definecolor{journalLinks}{rgb}{0.6, 0, 0}
\usepackage{amssymb}
\usepackage{amsmath,color}
\usepackage{epsfig}
\usepackage{graphicx,epsf,epsfig}
\usepackage{footnote}
\usepackage{multirow}
\usepackage{float}
\usepackage{enumitem}
\usepackage{color}
\def\slc#1{\setbox0=\hbox{$#1$}                  
    \dimen0=\wd0                                 
    \setbox1=\hbox{/} \dimen1=\wd1               
    \ifdim\dimen0>\dimen1                        
       \rlap{\hbox to \dimen0{\hfil/\hfil}}      
       #1                                        
    \else                                        
       \rlap{\hbox to \dimen1{\hfil$#1$\hfil}}   
       /                                         
    \fi}

\def\be{\begin{equation}}
\def\ee{\end{equation}}
\def\gs{\mathrel{
   \rlap{\raise 0.511ex \hbox{$>$}}{\lower 0.511ex \hbox{$\sim$}}}}
\def\ls{\mathrel{
   \rlap{\raise 0.511ex \hbox{$<$}}{\lower 0.511ex \hbox{$\sim$}}}}

\newcommand{\ba}{\begin{array}{c}}
\newcommand{\baz}{\begin{array}{cc}}
\newcommand{\barrr}{\begin{array}{rrr}}
\newcommand{\bad}{\begin{array}{ccc}}
\newcommand{\bav}{\begin{array}{cccc}}
\newcommand{\baf}{\begin{array}{ccccc}}
\newcommand{\bea}{\begin{equation} \begin{array}{c}}
\newcommand{\eea}{\end{array} \end{equation}}
\newcommand{\ea}{\end{array}}

\usepackage[normalem]{ulem}

\def\21{$\mathrm{SU(2)_L \otimes U(1)_Y}$ }

\newcommand {\ignore}[1]{}

\newcommand{\vt}{\vert}
\newcommand{\nn}{\nonumber}

\newcommand{\mc}{\mathcal}
\allowdisplaybreaks \allowdisplaybreaks[2]

\newcommand{\AddrCECYT}{Centro de Estudios Cient\'ificos y Tecnol\'ogicos No 16, Instituto Polit\'ecnico Nacional, Pachuca: Ciudad del Conocimiento y la Cultura, Carretera Pachuca Actopan km 1+500, San Agust\'in Tlaxiaca, Hidalgo, M\'exico.\\ 
}

\newcommand{\AddrESFM}{Escuela Superior de Física y Matemáticas, Instituto Politécnico Nacional, Edificio 9, Mexico City 07738, Mexico}

\begin{document}
\title{A lepton model with nearly Cobimaximal mixing} 
%
\author{Juan Carlos G\'omez-Izquierdo}
\email{cizquierdo@ipn.mx}
\affiliation{\AddrCECYT}

\author{Asahel Enrique Pozas Ramírez}
\email{apozasr1700@alumno.ipn.mx}
\affiliation{\AddrESFM}

%

%

\date{\bf \today} 

\begin{abstract}\vspace{2cm}
Cobimaximal mixing predicts $\pi/4$ and $3\pi/2$ for the atmospheric angle and the Dirac CP-violating phase, respectively. These values are in tension with the neutrino globals fits. If this pattern was behind the lepton mixings, then it would have to be broken. In that case, in this paper, we explore the $\mathbf{S}_{3}$ flavor symmetry within the $B-L$ gauge model where the aforementioned scheme comes from the neutrino sector but the charged lepton contribution breaks the well known predictions so the mixing observables as well as the $m_{ee}$ mass can be accommodated quite well according to the  available data. Notably, the predicted regions for the Dirac CP-violating phase would allow us to test the model in future experiments.
\end{abstract}

\begin{flushright}
CECyT 16-24-I
\end{flushright}

\maketitle
%

\section{Introduction}

The experimental evidence has shown neutrinos have mass, there is more matter than antimatter and dark matter undoubtedly exists. These observations have motivated us to look for answers beyond the Standard Model (SM) since that it is not the last theory, even though its tremendous success, and extra ingredients should be added to explain the aforementioned subtle issues.

The Higgs physics, which is currently in developing, will reveal many properties of the discovered Higgs boson at LHC in 2012~\cite{ATLAS:2012yve,CMS:2012qbp}. All of this will play an important role to have a complete understanding of the Higgs sector that might be linked with fundamental problems~\cite{Salam:2022izo, Peskin:2023sed}. In this line of thought, the Higgs and neutrino sector could be related indirectly by the type I see-saw mechanism~\cite{Minkowski:1977sc,Yanagida:1979as,GellMann:1980vs,Mohapatra:1980yp,Mohapatra:1979ia,Schechter:1980gr,Schechter:1981cv} which utilizes a Dirac mass that comes from the Higgs mechanism. An another connection might have to do with the peculiar pattern that exhibits the PMNS mixing matrix~\cite{Maki:1962mu, Pontecorvo:1967fh}, in the scheme of three species of neutrinos, large values in its entries one can find as well as one can observe that $\vert \mathbf{U}_{\mu i}\vert\approx \vert \mathbf{U}_{\tau i}\vert$ ($i=1,2,3$)~\cite{deSalas:2020pgw,Esteban:2020cvm}.  This latter fact seems to indicate us the presence of a symmetry, for instance $\mu-\tau$ reflection symmetry~\cite{Xing:2022uax} that can be identified with Cobimaximal pattern~\cite{Fukuura:1999ze,Miura:2000sx,Ma:2002ce,Grimus:2003yn,Chen:2014wxa,Ma:2015fpa,Joshipura:2015dsa,Li:2015rtz,He:2015xha,Chen:2015siy,Ma:2016nkf,Damanik:2017jar,Ma:2017trv,Grimus:2017itg,CarcamoHernandez:2017owh,CarcamoHernandez:2018hst,Ma:2019iwj} whose predictions on the Dirac phase, atmospheric and reactor angles are $3\pi/2$, $\pi/4$ and $\theta_{13}\neq 0$, respectively. From the model building point of view, this kind of pattern may be achieved by means of non-abelian discrete groups~\cite{Ishimori:2010au, Grimus:2011fk, Altarelli:2012bn, Altarelli:2012ss, King:2013eh, King:2015aea,Fonseca:2014lfa,Chauhan:2022gkz} but  one may require extra scalar, fermion or exotic fields. Up to now, there is one Higgs doublet then a well motivated framework is to include many Higgses, that transform in a non-trivial way under the discrete group, to shape the fermion mass matrices and therefore their mixings. This kind of extensions possess a rich phenomenology~\cite{Flores:1982pr,Barroso:2005da,Mantry:2007ar,Grimus:2008nb,Ferreira:2008zy,Botella:2009pq,Ferreira:2010xe,Blechman:2010cs,Diaz-Cruz:2014pla} such that flavored leptogenesis is a viable way to explain the excess of matter over antimatter. These theoretical ideas can be tested in the current or future experiments~\cite{Gehrlein:2022nss}.

Once the neutrino masses and mixing angles are added to the well established quark ones, the flavor puzzle~\cite{Feruglio:2015jfa,Abbas:2023ivi,Nilles:2023shk} is awaiting for a compelling solution. In this sense, the flavor symmetries might be a route to find out the underlaying physics in the flavor sector and a lot of progress has been made in this direction such that appealing patterns~\cite{Ishimori:2010au, Grimus:2011fk, Altarelli:2012bn, Altarelli:2012ss, King:2013eh, King:2015aea,Fonseca:2014lfa,Chauhan:2022gkz} have been proposed to accommodate the available experimental data. Nonetheless, there is too much work to realize in order to embed the ideal discrete symmetry in the adequate theoretical framework to describe all the observables.

With this in mind, we will recover the well known $\mathbf{S}_{3}$~\cite{Ishimori:2010au} symmetry, the permutation of the three objects, which has been implemented in different frameworks~\cite{Kubo:2005sr,Mondragon:2007af,Mondragon:2007nk,Barranco:2010we,Meloni:2010aw,Kubo:2012ty,Canales:2012dr,Canales:2013cga,Hernandez:2014lpa,Hernandez:2014vta,Hernandez:2015dga,Hernandez:2015zeh,Hernandez:2015hrt,Hernandez:2013hea,Ma:2014qra,Gomez-Izquierdo:2017rxi, Ge:2018ofp,Dicus:2010iq,Das:2015sca,Pramanick:2016mdp,Garces:2018nar} for studying masses and mixings. There are still strong motivations to study on this symmetry, for instance, the hierarchy among the fermion masses is understood by its irreducible representations and it may come from string theories~\cite{Fischer:2013qza, Nilles:2023shk}. On the other hand, the $B-L$ gauge model~\cite{Davidson:1978pi, Marshak:1979fm,Wetterich:1981bx, Masiero:1982fi,Buchmuller:1991ce} is a minimal extension of the SM where the type I see-saw mechanism~\cite{Minkowski:1977sc,GellMann:1980vs,Yanagida:1979as,Mohapatra:1980yp,Mohapatra:1979ia,Schechter:1980gr,Schechter:1981cv} is realized by  the inclusion of three right-handed neutrinos ($N$), which are required by the  SM$\otimes \mathbf{U}(1)_{B-L}$ gauge group to  cancel anomalies. These neutral fermions acquire their mass similarly to the quarks and charged leptons do in the SM, this is, a Higgs mechanism is realized by the presence of $B-L$ charged singlet field ($\phi$) which develops a vacuum expectation value (vev). Eventually, the light  neutrinos, that take place in the oscillations, get tiny masses. In addition to neutrino masses and mixings, a rich phenomenology as leptogenesis, dark matter and inflation can be found in the $B-L$ model~\cite{Khalil:2006yi, Emam:2007dy, Khalil:2008ps,Higaki:2014dwa,Rodejohann:2015lca}.

Motivated by the above issues, we embed the $\mathbf{S}_{3}$ symmetry in the $B-L$ model so that the family number is extended to the scalar sector. Then, three Higgs doublets and singlets (charged under $B-L$) are included to generate the mixings. Nonetheless, the scalar ($(H_{1},H_{2})\sim \mathbf{2}$, $H_{3}\sim \mathbf{1_{S}}$) and lepton ($l_{1} \sim \mathbf{1_{S}}$, $(l_{2},l_{3})\sim \mathbf{2}$) families are not treated in the same footing under the irreducible representation of the $\mathbf{S}_{3}$ symmetry. The main purpose to realize it, in the scalar sector, is to use the results on the allowed alignments that scalar potential with three Higgs doublets together with $\mathbf{S}_{3}$ permits~\cite{Beltran:2009zz, Emmanuel-Costa:2016vej, Kuncinas:2020wrn, Khater:2021wcx, Kuncinas:2023ycz}. Although many scalar fields have been added, the number of Yukawa couplings are reduced substantially by imposing an extra  symmetry ($\mathbf{Z}_{2}$). Along with this, the assignation
makes it possible to identify the 
Cobimaximal pattern, in the effective neutrino sector, by assuming real Dirac and Majorana Yukawa couplings. Whereas the charged lepton mass matrix ends up being complex due to the charged Yukawa couplings are assumed to be complex. In this benchmark, we calculate explicitly the PMNS matrix that is mainly controlled by the Cobimaximal one while the charged lepton breaks the well known predictions on the atmospheric angle ($\theta_{23}$) and Dirac CP-violating phase ($\delta_{CP}$). Notably, the PMNS mixing matrix, and consequently,  the aforementioned observables depend on two free parameters, this feature makes the model completely different other flavored ones~\cite{Vien:2021uxw, Vien:2021eog,Vien:2021bsv,Vien:2021pqg,Vien:2022cza}. An analytical and numerical study on  the atmospheric angle  and Dirac CP-violating phase allows to constrain the allowed region of values for the free parameters. In consequence, our findings are as follows: the former observable can be on the upper or lower octant for the normal and inverted hierarchy. Besides this, for the $\delta_{CP}$ phase, the predicted region is more constrained for the inverted hierarchy in comparison to the normal case, these facts would allow to test the model by confronting our results with future experiments~\cite{JUNO:2015zny,DUNE:2020ypp,Hyper-Kamiokande:2018ofw}. Taking into account the $\vert m_{ee}\vert $ effective neutrino mass, this is controlled totally by the reactor and solar angle, so this is consistent with the experimental data.

The letter is organized as follows. The $B-L$ model is revisited briefly in Section II. The $\mathbf{S}_{3}$ discrete symmetry is embedded in the $B-L$ model, the masses and mixings are obtained in Section III. In Section IV, we perform explicitly the PMNS mixing matrix as well as the deviation to the atmospheric angle and the Dirac CP-violating phase. The findings are highlighted through some scattered plots. In addition, in Section V, the $\vert m_{ee}\vert$ 
effective neutrino is calculated numerically. Finally, we end up giving a concluding remarks in Section VI.

\section{B-L Model}
One of the gauge models, that takes into account the missing neutrino mass term, is the well known Baryon minus Lepton (B-L) one~\cite{Davidson:1978pi, Marshak:1979fm,Wetterich:1981bx, Masiero:1982fi,Buchmuller:1991ce} where the inclusion of three right-handed neutrinos ($N$) are required by the gauge group. These neutral fermions acquire their mass similarly to the quarks and charged leptons do in the SM framework, this is, a Higgs  mechanism is realized by the presence of singlet field ($\phi$) which develops a vacuum expectation value (vev). Eventually, the light (active) neutrinos, that take place in the oscillations, get tiny mass due to the type I see-saw mechanism~\cite{Minkowski:1977sc,GellMann:1980vs,Yanagida:1979as,Mohapatra:1980yp,Mohapatra:1979ia,Schechter:1980gr,Schechter:1981cv}.

In the minimal version, under the  $\mathbf{SU(3)}_{c}\otimes \mathbf{SU(2)}_{L}\otimes
\mathbf{U(1)}_{Y}$ gauge group, the matter content has the usual quantum numbers. In table \ref{TB1}, we just show the quantum numbers under the $ \mathbf{U(1)}_{B-L}$ group.
\begin{table}[ht]
	\begin{center}
	\begin{tabular}{|c|c|c|c|c|c|c|c|c|c|c|c|c|c|c|c}
	\hline \hline
Matter & Quarks & Leptons & Higgs & $\phi$	\\ \hline
B-L &	$1/3$ & $1$ & $0$ & $-2$ \\\hline \hline
\end{tabular}\caption{Minimal {\bf $B-L$} model}\label{TB1}
\end{center}
\end{table}

Hence, a part from the SM Lagrangian, one has to add extra terms. This is
\begin{equation}
	\mathcal{L}_{BL}=\mathcal{L}_{SM}-y^{D}\bar{L}\tilde{H}N-\frac{1}{2}y^{N}\bar{N}^{c}\phi N-V\left(H,\phi\right)\label{EQ1}
\end{equation}
with
\begin{equation}
	V\left(H,\phi \right)=\mu^{2}_{BL}\phi^{\dagger} \phi+\frac{\lambda_{BL}}{2}\left(\phi^{\dagger} \phi\right)^{2}-\lambda_{H \phi}\left(H^{\dagger} H \right)\left(\phi^{\dagger} \phi \right).\label{EQ2}
\end{equation}
where $\tilde{H}_{i}=i\sigma_{2}H^{\ast}_{i}$ and $\sigma_{2}$ being the second Pauli matrix.

In this framework, there are two scales of the spontaneous symmetry breaking, the first one is associated to the $\mathbf{U(1)}_{B-L}$ group and
the breaking scale is larger than the electroweak scale, $\phi_{0}\gg
v$. This latter associates to $\mathbf{SU(2)}_{L}$.
\begin{equation}
	\langle H\rangle=\frac{1}{\sqrt{2}}\begin{pmatrix}
		0	\\ v
	\end{pmatrix}, \quad \langle \phi\rangle=\frac{\phi_{0}}{\sqrt{2}}.\label{EQ3}
\end{equation}

In the standard basis, the fermion mass term is given by
\begin{align}
	-\mathcal{L}_{Y}=\bar{\ell}_{i L} \left( {\bf M}_{\ell}\right)_{ij}\ell_{j R}
	+\dfrac{1}{2}\bar{\nu}_{i L}\left({\bf M}_{\nu}\right)_{ij}\nu^{c}_{j L }+\dfrac{1}{2}\bar{N}^{c}_{i}\left({\bf M}_{R}\right)_{ij}N_{j }+h.c.\label{EQ5}
\end{align}
where the type I see-saw was realized.

In general, the above mass matrices are complex, and they do not possess any pattern which help us find out the mixing matrices that take place in PMNS mixings one. Therefore, we have the need to use discrete symmetries as guidance to explain the peculiar patterns. 
\section{Lepton sector}


As we already commented, the scalar sector of the $B-L$ model is augmented such that three Higgs doublets ($H_{i}$) and singlets ($\phi_{i}$) are considered. Then, for leptons the assignation under the flavor symmetry is given as follows: the first family is put in a $\mathbf{1}_{S}$ singlet; the second and third families live in a $\mathbf{2}$ doublet. On the other side, the scalar sector is assigned as the usual studios have realized it, this is, the first and second family are put together within a $\mathbf{2}$ and the third one belongs to $\mathbf{1}_{S}$.
The main reason to keep the typical assignation in the scalar sector is to take into account 
the results on the $\mathbf{S}_{3}$ scalar potential with three Higgs where the aforementioned assignation was utilized and an exhaustive classification about the vev's alignments was released in~\cite{Emmanuel-Costa:2016vej, Kuncinas:2020wrn, Khater:2021wcx, Kuncinas:2023ycz}.  

On the other hand, some Yukawa couplings are forbidden by imposing the $\mathbf{Z}_{2}$ symmetry. In this manner, the Dirac and charged lepton mass matrices end up being almost diagonal, then the right-handed neutrino mass matrix will provide the mixings. 

The table \ref{TAB2} shows explicitly the full assignment for the matter in the current model. 
\begin{table}[ht]
	\begin{center}
		\begin{tabular}{|c|c|c|c|c|c|c|c|c|c|c|c|c|c|c|}
			\hline \hline	
			{\footnotesize Matter} & {\footnotesize $H_{I}, L_{J}, e_{J R}, N_{J}$} & {\footnotesize $L_{1}, e_{1 R}, N_{1}$} & {\footnotesize $H_{3}, \phi_{3}$} & {\footnotesize $\phi_{I}$}   \\ \hline
			{\footnotesize \bf $\mathbf{S}_{3}$} &  {\footnotesize \bf $2$} & {\footnotesize \bf $1_{S}$}   & {\footnotesize \bf $1_{S}$} & {\footnotesize \bf $2$} \\ \hline
			{\footnotesize \bf $\mathbf{Z}_{2}$} & {\footnotesize $1$} & {\footnotesize $-1$}  &  {\footnotesize $1$} & {\footnotesize $-1$} \\ \hline \hline
		\end{tabular}\caption{Flavored $B-L$ model. Here, $I=1,2$ and $J=2,3$.}\label{TAB2}
	\end{center}
\end{table}

The Yukawa mass term that respects the ${\bf S}_{3}\otimes {\bf Z}_{2}$ flavor symmetry and the gauge group, is given by
\begin{align}
-\mathcal{L}_{Y}&=y^{e}_{1}\bar{L}_{1}H_{3}e_{1 R}+y^{e}_{2}\left[(\bar{L}_{2}H_{2}+\bar{L}_{3}H_{1})e_{2 R}+(\bar{L}_{2}H_{1}-\bar{L}_{3}H_{2})e_{3 R} \right]+y^{e}_{3}\left[\bar{L}_{2}H_{3}e_{2 R}+\bar{L}_{3}H_{3}e_{3 R}\right]+y^{D}_{1}\bar{L}_{1}\tilde{H}_{3}N_{1}\nn\\&
+y^{D}_{1}\bar{L}_{1}\tilde{H}_{3}N_{1}+ y^{N}_{2}\left[(\bar{L}_{2} \tilde{H}_{2}+\bar{L}_{3} \tilde{H}_{1})N_{2}+(\bar{L}_{2}\tilde{H}_{1}-\bar{L}_{3} \tilde{H}_{2})N_{3 } \right]
+y^{N}_{1}\bar{N}^{c}_{1}\phi_{3}N_{1}+y^{N}_{2}\bigg[\bar{N}^{c}_{1}\left(\phi_{1 }N_{2}+\phi_{2}N_{3}\right)\nn\\&+\left(\bar{N}^{c}_{2}\phi_{1}+\bar{N}^{c}_{3}\phi_{2}\right)N_{1}\bigg]+y^{N}_{3}\left[\bar{N}^{c}_{2}\phi_{3 }N_{2}+\bar{N}^{c}_{3}\phi_{3}N_{3}\right]+h.c.\label{EQ4}
\end{align}

From Eq.(\ref{EQ4}), the mass matrices have the following form
\begin{align}
{\bf M}_{e}&=\begin{pmatrix}
y^{e}_{1}\langle H_{3}\rangle & 0 & 0 \\ 
0 & y^{e}_{3}\langle H_{3}\rangle+y^{e}_{2}\langle H_{2}\rangle & y^{e}_{2}\langle H_{1}\rangle \\ 
0 & y^{e}_{2}\langle H_{1}\rangle & y^{e}_{3}\langle H_{3}\rangle-y^{e}_{2}\langle H_{2}\rangle \end{pmatrix};\nn\\
{\bf M}_{D}&=\begin{pmatrix}
	y^{D}_{1}\langle \tilde{H}_{3}\rangle & 0 & 0 \\ 
	0 & y^{D}_{3}\langle \tilde{H}_{3}\rangle+y^{D}_{2}\langle \tilde{H}_{2}\rangle & y^{D}_{2}\langle \tilde{H}_{1}\rangle \\ 
	0 & y^{D}_{2}\langle \tilde{H}_{1}\rangle & y^{D}_{3}\langle \tilde{H}_{3}\rangle-y^{D}_{2}\langle \tilde{H}_{2}\rangle \end{pmatrix};\nn\\
 {\bf M}_{R}&=\begin{pmatrix}
y^{N}_{1} \langle \phi_{3}\rangle & y^{N}_{2} \langle \phi_{1}\rangle & y^{N}_{2} \langle \phi_{2}\rangle \\ 
y^{N}_{2} \langle \phi_{1}\rangle & y^{N}_{3}\langle \phi_{3}\rangle & 0 \\ 
y^{N}_{2} \langle \phi_{2}\rangle & 0 & y^{N}_{3}\langle \phi_{3}\rangle
	\end{pmatrix}.\label{EQ6} 
\end{align}

Before finishing this section, we would like to add a comment about the scalar potential in the scenario of three Higgs doublets with $\mathbf{S}_{3}$ symmetry (3HD-S3). As it is well known, the scalar potential study becomes mandatory to get a realistic scenario that intents to explain the mixing patterns. In this direction, works on the minimization of the scalar potential, vev's alignments, mass spectrum and phenomenology can be found in~\cite{Pakvasa:1977in, Kubo:2004ps, EmmanuelCosta:2007zz, Beltran:2009zz, Teshima:2012cg, Das:2014fea, Barradas-Guevara:2014yoa, Gomez-Bock:2021uyu}. Besides this, exhaustive studios have released significant results on the minimization of the scalar potential and alignments that are allowed~\cite{Emmanuel-Costa:2016vej, Kuncinas:2020wrn, Khater:2021wcx, Kuncinas:2023ycz} in the 3HD-S3 framework. Remarkably, those have provided a larger list of alignments which must have great impact on the gauge sector, fermion and scalar masses. Having commented that, in our model, we will take advantage of the previous results,  so it is not necessary to analyze again the potential. The reason is the following: the full potential is given by $V_{B-L}=V(H)+V(\phi)+V(H,\phi)$, then
we make a strong assumption, this is, the Higgses and singlets do not interact between them so one can treat them separately in consequence  $V(H)$ and $V(\phi)$ share the same structure under the flavor symmetry as a result of this we can use the analysis realized in~\cite{Emmanuel-Costa:2016vej, Kuncinas:2020wrn, Khater:2021wcx, Kuncinas:2023ycz}. A partial study on the $V_{B-L}$ potential was carried out in \cite{Gomez-Izquierdo:2018jrx}.

\subsection{Masses and mixings}
One of the allowed alignments for the vev's is given by 
$\langle H_{1}\rangle=v_{1}$, $\langle H_{2}\rangle=iv_{2}$ and $\langle H_{3}\rangle=v_{3}$~\cite{Emmanuel-Costa:2016vej, Kuncinas:2020wrn, Khater:2021wcx, Kuncinas:2023ycz}. 
Hence, we will utilize those in the Dirac neutrinos and charged lepton. It is worth mention that it has not been yet explored in the quark sector and this is a work in progress. In the Majorana sector, the right-handed neutrinos get their mass through the $\phi$ scalars so that we will also consider the alignment
$\langle \phi_{2}\rangle=\langle \phi_{1}\rangle$ which is also a solution of the scalar potential as was shown in~\cite{ Beltran:2009zz}.

Having chosen the alignments, we have
\begin{equation}
		{\bf M}_{e}=\begin{pmatrix}
			a_{e} & 0 & 0 \\ 
			0 & c_{e}+i b_{e} & d_{e} \\ 
			0 & d_{e} & c_{e}-i b_{e}
		\end{pmatrix},\qquad 	{\bf M}_{D}=\begin{pmatrix}
		a_{D} & 0 & 0 \\ 
		0 & c_{D}-i b_{D} & d_{D} \\ 
		0 & d_{D} & c_{D}+i b_{D}
	\end{pmatrix},\qquad {\bf M}_{R}=\begin{pmatrix}
	a_{R} & b_{R} & b_{R} \\ 
	b_{R} & c_{R} & 0 \\ 
	b_{R} & 0 & c_{R}
\end{pmatrix}
	\end{equation}
where
\begin{align}
&a_{e}=y^{e}_{1} v_{3},\quad c_{e}=y^{e}_{3} v_{3},\quad b_{e}=y^{e}_{2} v_{2},\quad d_{e}=y^{e}_{2} v_{1}, \nn\\&
a_{D}=y^{D}_{1}v_{3},\quad c_{D}=y^{D}_{3}{v}_{3},\quad b_{D}=y^{D}_{2}v_{2},\quad d_{D}=y^{D}_{2} v_{1},\nn\\ 	
&a_{R}=y^{N}_{1} \langle \phi_{3}\rangle,\quad b_{R}=y^{N}_{2} \langle \phi_{1}\rangle,\quad c_{R}=y^{N}_{3}\langle \phi_{3}\rangle .
\label{eq9}
\end{align}
\subsubsection{Charged lepton mixings}

In this brief section, let us obtain the mixing matrix that takes place in the PMNS one. To do this, we assume that the charged Yukawa couplings are complex. Then, $\mathbf{M}_{e}$ mass matrix is written as
\begin{equation}
	{\bf M}_{e}=\begin{pmatrix}
		a_{e} & 0 & 0 \\ 
		0 & A_{e} & d_{e} \\ 
		0 & d_{e} & B_{e}
	\end{pmatrix},
\end{equation}
with $A_{e}=c_{e}+ib_{e}$ and $B_{e}=c_{e}-ib_{e}$. The above matrix is diagonalized by $\mathbf{U}_{eL}=\mathbf{P}_{e}\mathbf{O}_{e}$ and $\mathbf{U}_{eR}=\mathbf{P}^{\dagger}_{e}\mathbf{O}_{e}$ such that $\hat{{\bf M}}_{e}=\textrm{Diag.}\left(m_{e}, m_{\mu}, m_{\tau}\right)=\mathbf{U}^{\dagger}_{e L} \mathbf{M}_{e} \mathbf{U}_{e R}$. Explicitly, we obtain
\begin{equation}
		{\bf O}_{e}=\begin{pmatrix}
			1 & 0 & 0 \\ 
			0 & \cos{\theta}_{e} & \sin{\theta}_{e} \\ 
			0 & -\sin{\theta}_{e} & \cos{\theta}_{e}
		\end{pmatrix},\qquad   {\bf P}_{e}=\textrm{Diag.}\left(e^{i\eta_{e}}, e^{i\eta_{\mu}},e^{i\eta_{\tau}}\right).\label{EL2}
	\end{equation}
Along with this, $\eta_{e}=\textrm{arg.}(a_{e})/2$, $\eta_{\mu}=\textrm{arg.}(A_{e})/2$,  $\eta_{\tau}=\textrm{arg.}(B_{e})/2$ and $\eta_{\mu}+\eta_{\tau}=\textrm{arg.}(b_{e})$. At the same time, we have
\begin{equation}
\cos{\theta_{e}}=\sqrt{\frac{m_{\tau}-\vert A_{e}\vert}{m_{\tau}-m_{\mu}}},\qquad \sin{\theta_{e}}=\sqrt{\frac{\vert A_{e}\vert- m_{\mu}}{m_{\tau}-m_{\mu}}}.\label{EL3}
\end{equation}

Notice that there is a free parameter, $\vt A_{e}\vt$, which has to satisfy the following constraint $m_{\tau}>\vt A_{e}\vt> m_{\mu}$.

\subsubsection{Neutrino mixings}
According to the previous section, we considered $\langle \phi_{1}\rangle=\langle \phi_{2}\rangle$. For this reason, in the Majorana sector, the free parameters are reduced in $\mathbf{M}_{R}$, and the inverse matrix is parametrized as
\begin{equation}
 	{\bf M}^{-1}_{R}=\begin{pmatrix}
	\mc{X}& -\mc{Y} & -\mc{Y} \\ 
	-\mc{Y} & \mc{W} & \mc{Z} \\ 
	-\mc{Y} & \mc{Z} & \mc{W} 
\end{pmatrix}, 
\end{equation}
where we have defined $\mc{X}=c^{2}_{R}/\vert \mathbf{M}_{R}\vert$, $\mc{Y}=b_{R}c_{R}/\vert \mathbf{M}_{R}\vert$, $\mc{W}=\left(a_{R}c_{R}-b^{2}_{R}\right)/\vert \mathbf{M}_{R}\vert$ and $\mc{Z}=b^{2}_{R}/\vert \mathbf{M}_{R}\vert$. Here, 
$\vert \mathbf{M}_{R}\vert$ denotes the determinant of $\mathbf{M}_{R}$. With all of this, the effective neutrino mass matrix, ${\bf M}_{\nu}=-{\bf M}_{D} {\bf M}^{-1}_{R} {\bf M}^{T}_{D}$, is given by
{\footnotesize
\begin{equation}
		{\bf M}_{\nu}=\begin{pmatrix}
			\mc{X} a^{2}_{D} & -\mc{Y}a_{D}\left[c_{D}+ d_{D}-ib_{D}\right]  & -\mc{Y}a_{D} \left[c_{D}+ d_{D}+ib_{D}\right] \\ 
			-\mc{Y}a_{D}\left[c_{D}+ d_{D}-ib_{D}\right] & \mc{W}\left[\left(c_{D}-ib_{D}\right)^{2}+d^{2}_{D}\right]+2\mc{Z}d_{D}\left[c_{D}-ib_{D}\right]  &  2\mc{W}d_{D}c_{D}+ \mc{Z}\left[b^{2}_{D}+c^{2}_{D}+d^{2}_{D}\right]\\ 
			-\mc{Y}a_{D} \left[c_{D}+ d_{D}+ib_{D}\right] & 2\mc{W}d_{D}c_{D}+ \mc{Z}\left[b^{2}_{D}+c^{2}_{D}+d^{2}_{D}\right] & \mc{W}\left[\left(c_{D}+ib_{D}\right)^{2}+d^{2}_{D}\right]+2\mc{Z}d_{D}\left[c_{D}+ib_{D}\right]
\end{pmatrix},			
\end{equation}		}	
which can be parametrized as
\begin{equation}
{\bf M}_{\nu}\equiv \begin{pmatrix}
		A_{\nu} & B_{\nu}  &  B^{\ast}_{\nu} \\ 
		B_{\nu} & C^{\ast}_{\nu} & D_{\nu}  \\ 
		B^{\ast}_{\nu} & D_{\nu}  & C_{\nu}
		\end{pmatrix}.
\end{equation}
We ought to stress that one can identify the above mass matrix with the Cobimaximal one if the Dirac Yukawa couplings and the vev's values are real. In this benchmark, the PMNS mixings are governed mainly by the neutrino sector. It is well known that
$\mathbf{M}_{\nu}$ is diagonalized by $\mathbf{U}_{\nu}=\mathbf{U}_{\alpha}\mathbf{O}_{23}\mathbf{O}_{13}\mathbf{O}_{12}\mathbf{U}_{\beta}$ such that $\hat{\mathbf{M}}_{\nu}=\textrm{Diag.}\left(\vert m_{1}\vert, \vert m_{2}\vert, \vert m_{3}\vert\right)= \mathbf{U}^{\dagger}_{\nu}\mathbf{M}_{\nu}\mathbf{U}^{\ast}_{\nu}$ where the diagonal mass matrices are given as $\mathbf{U}_{\alpha}=\textrm{Diag.}\left(e^{i \alpha_{1}}, e^{i \alpha_{2}}, e^{i \alpha_{3}}\right)$ and $\mathbf{U}_{\beta}=\textrm{Diag.}\left(1, e^{i \beta_{1}}, e^{i \beta_{2}}\right)$~\cite{Grimus:2003yn}. These stand for unphysical and Majorana phases, respectively. In addition, we have

\begin{equation}
\mathbf{O}_{23}=\begin{pmatrix}
1	& 0 & 0 \\
0	& \cos{\rho_{23}} & \sin{\rho_{23}} \\
0	& -\sin{\rho_{23}}  & \cos{\rho_{23}}
\end{pmatrix},\qquad 	\mathbf{O}_{13}=\begin{pmatrix}
\cos{\rho_{13}}	& 0 & \sin{\rho_{13}} e^{-i \delta} \\
0	& 1 & 0 \\
-\sin{\rho_{13}} e^{i \delta} 	& 0  & \cos{\rho_{13}}
\end{pmatrix},\qquad \mathbf{O}_{12}=\begin{pmatrix}
\cos{\rho_{12}}	& \sin{\rho_{12}} &  0 \\
-\sin{\rho_{12}}	& \cos{\rho_{12}} & 0 \\
0 	& 0  & 1
\end{pmatrix}.
\end{equation}

With $\rho_{23}=\pi/4$, $\delta=-\pi/2$, $\alpha_{1}=0=\alpha_{3}$ and $\alpha_{2}=\pi$; also,  $\beta_{1}=0$ and $\beta_{2}=\pi/2$. Thus, we reconstruct the neutrino mass matrix,  $\mathbf{M}_{\nu}=\mathbf{U}_{\nu}\hat{{\bf M}}_{\nu}\mathbf{U}^{T}_{\nu}$ , whose elements are  
\begin{eqnarray}
A_{\nu}&=&\vert m_{3}\vert \sin^{2}{\rho_{13}} + \cos^{2}{\rho_{13}}\left[\vert m_{1}\vert \cos^{2}{\rho_{12}}+\vert m_{2}\vert \sin^{2}{\rho_{12}}\right];\nn\\
B_{\nu}&=&\frac{\cos{\rho_{13}}}{\sqrt{2}}\left[\left(\vert m_{1}\vert-\vert m_{2}\vert\right)\cos{\rho_{12}}\sin{\rho_{12}}+i\sin{\rho_{13}}\left(\vert m_{3}\vert-\vert m_{1}\vert \cos^{2}{\rho_{12}}- \vert m_{2}\vert \sin^{2}{\rho_{12}}\right)\right];\nn\\
C_{\nu}&=&\frac{1}{2}\left[\vert m_{2}\vert\left(\cos{\rho_{12}}-i \sin{\rho_{13}}\sin{\rho_{12}}\right)^{2}+\vert m_{1}\vert\left(\sin{\rho_{12}}+i\cos{\rho_{12}}\sin{\rho_{13}}\right)^{2}-\vert m_{3}\vert\cos^{2}{\rho_{13}} \right];\nn\\
D_{\nu}&=& \frac{1}{2}\left[\vert m_{2}\vert\left(\cos^{2}{\rho_{12}}+ \sin^{2}{\rho_{13}}\sin^{2}{\rho_{12}}\right)+\vert m_{1}\vert\left(\sin^{2}{\rho_{12}}+\cos^{2}{\rho_{12}}\sin^{2}{\rho_{13}}\right)+\vert m_{3}\vert\cos^{2}{\rho_{13}}\right].
\end{eqnarray}
Finally, we obtain
\begin{equation}
\mathbf{U}_{\nu}=\begin{pmatrix}
\cos{\rho_{13}}\cos{\rho_{12}}	& \cos{\rho_{13}}\sin{\rho_{12}} & -\sin{\rho_{13}} \\
\frac{1}{\sqrt{2}}\left(\sin{\rho_{12}}-i\cos{\rho_{12}}\sin{\rho_{13}}\right)	& -\frac{1}{\sqrt{2}}\left(\cos{\rho_{12}}+i\sin{\rho_{12}}\sin{\rho_{13}}\right) & -i\frac{\cos{\rho_{13}}}{\sqrt{2}} \\
\frac{1}{\sqrt{2}}\left(\sin{\rho_{12}}+i\cos{\rho_{12}}\sin{\rho_{13}}\right)	& -\frac{1}{\sqrt{2}}\left(\cos{\rho_{12}}-i\sin{\rho_{12}}\sin{\rho_{13}}\right) & i\frac{\cos{\rho_{13}}}{\sqrt{2}}
\end{pmatrix}\label{Une}
\end{equation}

As one can realize, if $\rho_{13}=0$, it would imply the presence of $\mu \leftrightarrow \tau$ symmetry in $\mathbf{M}_{\nu}$ or a $2 \leftrightarrow 3$ exchange due to the charged lepton mass matrix is not diagonal. As a result, one expects a deviation to atmospheric angle and the Dirac phase.

\section{Results}
Once the relevant mixing matrices were performed, the PMNS one is defined as $\mathbf{U}=\mathbf{U}^{\dagger}_{e}\mathbf{U}_{\nu}$ and the theoretical matrix elements are given explicitly as
\begin{eqnarray}
\mathbf{U}_{11}&=& ({\bf {U}_{\nu}})_{11}~e^{-i\eta_{e}};\nn\\
\mathbf{U}_{12}&=& ({\bf {U}_{\nu}})_{12}~e^{-i\eta_{e}};\nn\\ 
\mathbf{U}_{13}&=& ({\bf {U}_{\nu}})_{13}~e^{-i\eta_{e}};\nn\\
\mathbf{U}_{21}&=& \left[\cos{\theta_{e}}({\bf {U}_{\nu}})_{21}- \sin{\theta_{e}}({\bf {U}_{\nu}})_{31}~e^{-i\eta_{\nu}}\right]e^{-i\eta_{\mu}};\nn\\
\mathbf{U}_{22}&=& \left[\cos{\theta_{e}}({\bf {U}_{\nu}})_{22}- \sin{\theta_{e}}({\bf {U}_{\nu}})_{32}~e^{-i\eta_{\nu}}\right]e^{-i\eta_{\mu}}
;\nn\\
\mathbf{U}_{23}&=&\left[\cos{\theta_{e}}({\bf {U}_{\nu}})_{23}- \sin{\theta_{e}}({\bf {U}_{\nu}})_{33}~e^{-i\eta_{\nu}} \right]e^{-i\eta_{\mu}};\nn\\
\mathbf{U}_{31}&=& \left[ \sin{\theta_{e}}({\bf {U}_{\nu}})_{21} +\cos{\theta_{e}}({\bf {U}_{\nu}})_{31}~e^{-i\eta_{\nu}} \right]e^{-i\eta_{\mu}};\nn\\
\mathbf{U}_{32}&=&\left[\sin{\theta_{e}}({\bf {U}_{\nu}})_{22} +\cos{\theta_{e}}({\bf {U}_{\nu}})_{32}~e^{-i\eta_{\nu}}\right]e^{-i\eta_{\mu}};\\
\mathbf{U}_{33}&=& \left[\sin{\theta_{e}}({\bf {U}_{\nu}})_{23} +\cos{\theta_{e}}({\bf {U}_{\nu}})_{33}~e^{-i\eta_{\nu}}\right]e^{-i\eta_{\mu}}.\label{UPMNS}
\end{eqnarray}
with $\eta_{\nu}\equiv \eta_{\tau}-\eta_{\mu}$ being a relative phase.  

In consequence, the expressions for the mixing angles are obtained by comparing our theoretical formula with the  
standard parametrization of the PMNS.
\begin{align}
\sin^{2}{\theta}_{13}&=\vert ({\bf U})_{13}\vert^{2} =\sin^{2}{\rho_{13}};\nn\\
\sin^{2}{\theta}_{23}&=\dfrac{\vert ({\bf U})_{23}\vert^{2}}{1-\vert {\bf U}_{13}\vert^{2}}=\frac{1}{2}\left[1+\sin{2\theta_{e}}\cos{\eta_{\nu}}\right];\nn\\
\sin^{2}{\theta_{12}}&= \dfrac{\vert ({\bf U})_{12}\vert^{2}}{1-\vert {\bf U}_{13}\vert^{2}}=\sin^{2}{\rho_{12}}
.\label{mixang}
\end{align} 

In addition, the Jarlskog invariant can be performed analytically. As one can verify straightforward, using Eqns.(\ref{Une})-(\ref{mixang}), we obtain 

\begin{eqnarray}
\sin{\delta_{CP}} &=& \frac{\textrm{Im}\left[(\mathbf{U})_{23}(\mathbf{U})^{\ast}_{13}(\mathbf{U})_{12}(\mathbf{U})^{\ast}_{22}\right]}{\frac{1}{8}\sin{2\theta_{12}}\sin{2\theta_{23}}\sin{2\theta_{13}}\cos{\theta_{13}}};\nn\\	
\sin{\delta_{CP}} &=&-\frac{1}{2}\frac{\cos{\rho_{12}}\sin{\rho_{12}}\sin{\rho_{13}} \cos^{2}{\rho_{13}}\cos{2\theta_{e}}}{\cos{\theta_{12}}\sin{\theta_{12}} \cos{\theta_{23}} \sin{\theta_{23}} \sin{\theta_{13}}\cos^{2}{\theta_{13}}};\nn\\
\sin{\delta_{CP}}&=&-\frac{\cos{2\theta_{e}}}{\sqrt{1-\sin^{2}{2\theta_{e}}\cos^{2}{\eta_{\nu}}}}.\label{IJE}
\end{eqnarray}

Having obtained the above expressions, some comments are added in order. The reactor and solar angles are controlled mainly by the neutrino sector. What is more these are associated directly to the $\rho_{13}$ and $\rho_{12}$ parameters, respectively. Also, the atmospheric angle is deviated from $\pi/4$ and the source of such deviation is the charged lepton sector as one would expect. Also, strictly speaking the magnitude of the PMNS depends on four free parameters namely $\rho_{12}$, $\rho_{13}$, $\theta_{e}$ (or $\vert A_{e}\vert$) and the $\eta_{\nu}$ relative phase. Nevertheless, as it was already remarked, $\rho_{13}=\theta_{13}$ and $\rho_{12}=\theta_{12}$ so we end up having two free parameters, this is, $\vert A_{e}\vert$ and $\eta_{\nu}$. These must be fixed by numerical study to get the PMNS values. Before making this, it is worthy pointed out extreme values for the mentioned parameters and their implications on the atmospheric angle and the Dirac CP-violating phase:

\begin{itemize}
\item If the $\eta_{\nu}$ effective phase was zero (or $\pi$), the atmospheric angle would lie on the upper (lower) octant and the $\delta_{CP}=-\pi/2$.
\item If the $\eta_{\nu}$ effective phase was $\pi/2$,  the atmospheric angle would be $\pi/4$ exactly and the $\delta_{CP}$ would be deviated from $-\pi/2$.
\item If $\vert A_{e}\vert\approxeq m_{\mu}$,  $\theta_{23}$ would be $\pi/4$ and $\delta_{CP}=-\pi/2$. Whereas, if  $\vert A_{e}\vert\approxeq m_{\tau}$ then the $\theta_{23}$ would be $\pi/4$ and $\delta_{CP}=\pi/2$. 
\item If $\vert A_{e}\vert=(m_{\mu}+m_{\tau})/2$ ($\theta_{e}=\pi/4$), then $\theta_{23}$ would lie on the upper or lower octant and $\delta_{CP}=\pi$ or $2\pi$.
\end{itemize}

In short, the atmospheric angle and the Dirac phase are totally different of $\pi/4$ and $3\pi/2$, respectively. Hence, scattered plots will be performed to find out the full set of allowed values for the free parameters ($\vert A_{e}\vert$ and $\eta_{\nu}$) that accommodates the observables up to $3\sigma$. To make this, the charged lepton masses~\cite{Xing:2022uax} will be considered like inputs as well as the atmospheric angle and Dirac phase. At the electroweak scale, we have
\begin{equation}
m_{e}=0.483 07 \pm 0.000 45~\textrm{MeV},\quad m_{\mu}=101.766 \pm 0.023~\textrm{MeV},  \quad m_{\tau}= 1728.56 \pm 0.28~\textrm{MeV}.
\end{equation}
Along with this, the experimental neutrino data are given at $3\sigma$ as follows~\cite{deSalas:2020pgw}
\begin{eqnarray}
\sin^{2}{\theta_{13}}&=& 0.02000-0.02405, \qquad \sin^{2}{\theta_{23}}= 0.434-0.610, \qquad \delta_{CP}/^{\circ}= 128-359;\qquad \textrm{Normal ordering}\nn\\
\sin^{2}{\theta_{13}}&=& 0.02018-0.02424,,\qquad \sin^{2}{\theta_{23}}= 0.433-0.608, \qquad \delta_{CP}/^{\circ}= 200-353.\qquad \textrm{Inverted  ordering}\label{OND}
\end{eqnarray}
In addition, $\sin^{2}{\theta_{12}}=0.271-0.369$ for the normal and inverted hierarchy. 
 
\begin{figure}[ht]
	\includegraphics[width=0.4\linewidth]{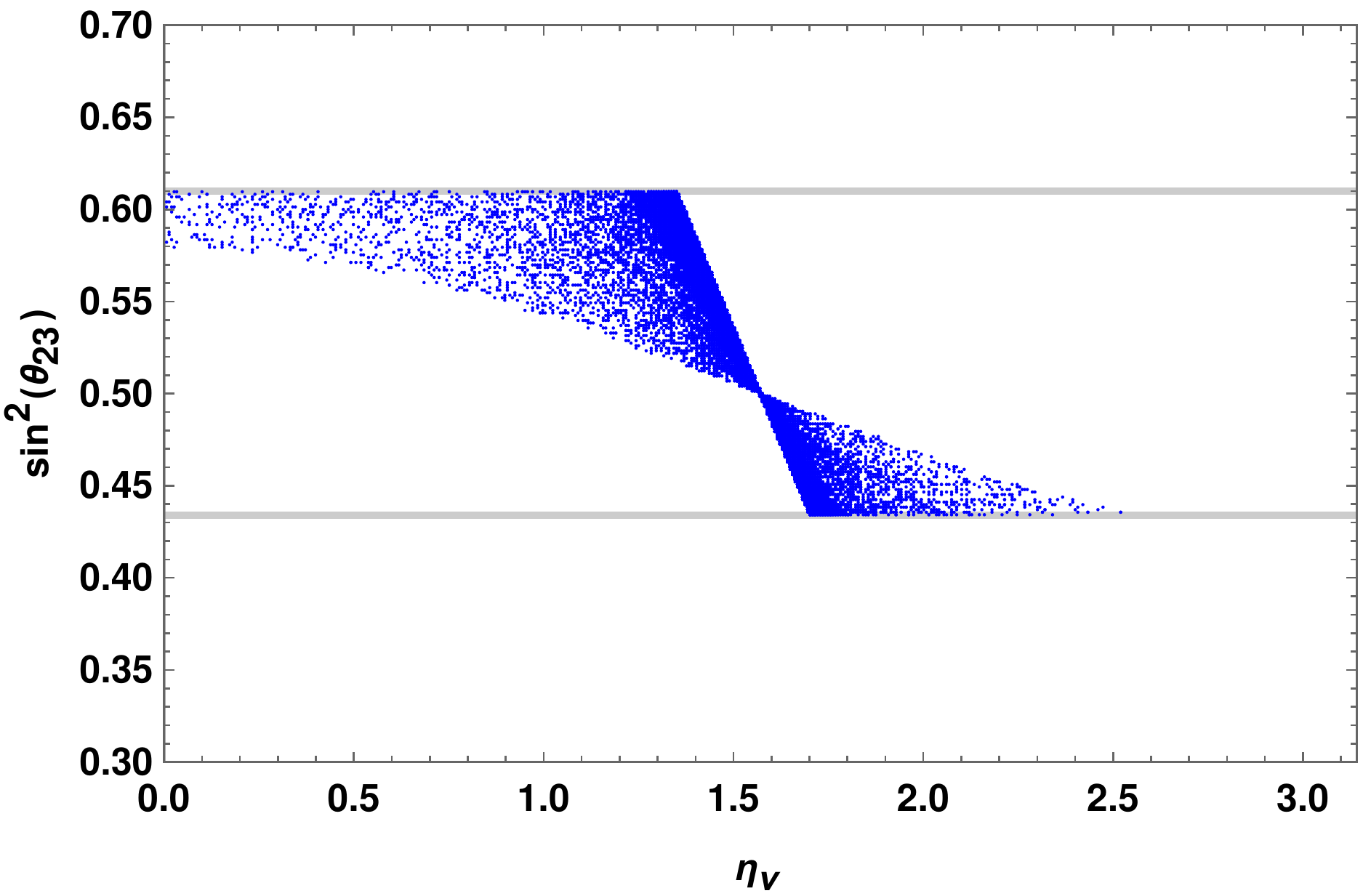}\hspace{5mm}\includegraphics[width=0.4\linewidth]{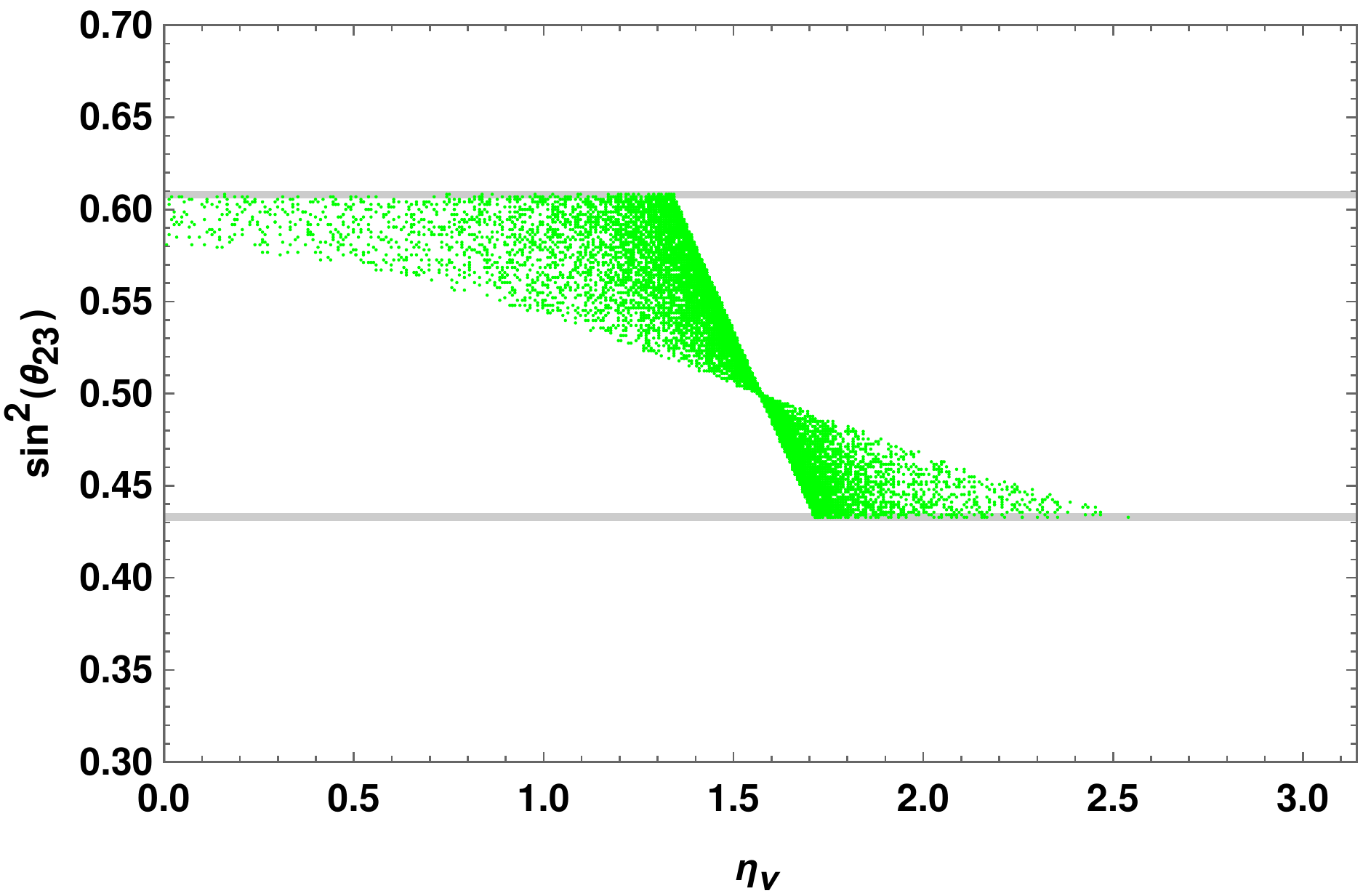}\hspace{5mm}
	\caption{\label{fig:s23vsa} 
Atmospheric angle as function of the $\eta_{\nu}$ relative phase for the normal and inverted hierarchy, respectively. The solid gray line stands for the experimental data at $3\sigma$.}
\end{figure}

From Eqns.(\ref{IJE}) and (\ref{mixang}), we vary arbitrarily the free parameters in their allowed range such that these fit the observables at $3\sigma$.
\begin{eqnarray}
\delta_{CP}\left(\vert A_{e}\vert (\theta_{e}), \eta_{\nu}\right)&=& \arcsin[-\frac{\cos{2\theta_{e}}}{\sqrt{1-\sin^{2}{2\theta_{e}}\cos^{2}{\eta_{\nu}}}}],\nn\\
\sin^{2}\theta_{23}\left(\vert A_{e}\vert (\theta_{e}), \eta_{\nu}\right)&=& \frac{1}{2}\left[1+\sin{2\theta_{e}}\cos{\eta_{\nu}}\right].
\end{eqnarray}
Having commented on that, let us show you the relevant plots. In Figure~\ref{fig:s23vsa}, we observe the set of values for the $\eta_{\nu}$ relative phase that fits the atmospheric angle. There are two regions where the angle lies on the upper and lower octant so there is a clear deviation around $\pi/4$. In addition, the allowed regions are similar for the normal and inverted hierarchy.

\begin{figure}[ht]
	\includegraphics[width=0.4\linewidth]{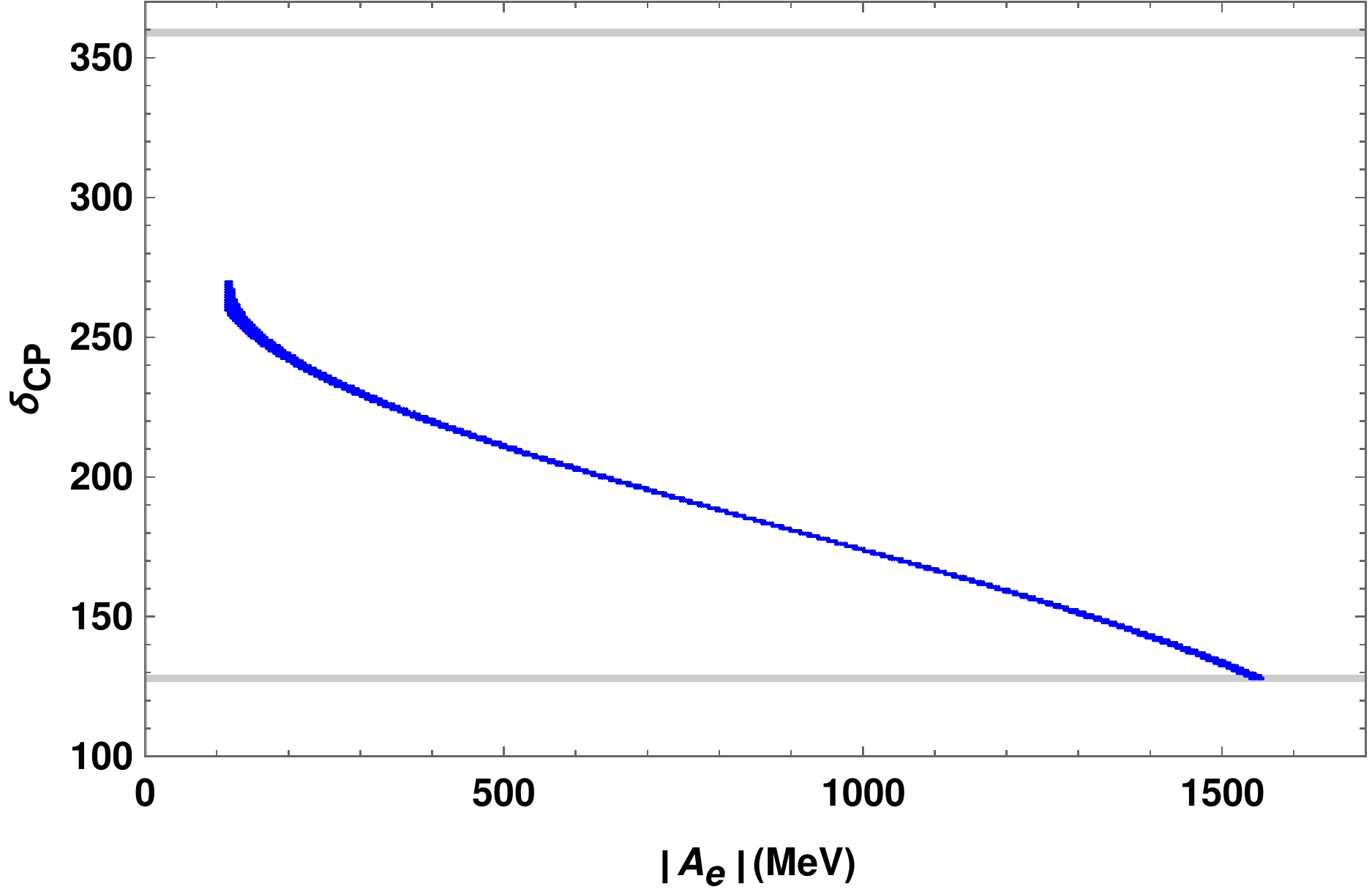}\hspace{5mm}\includegraphics[width=0.4\linewidth]{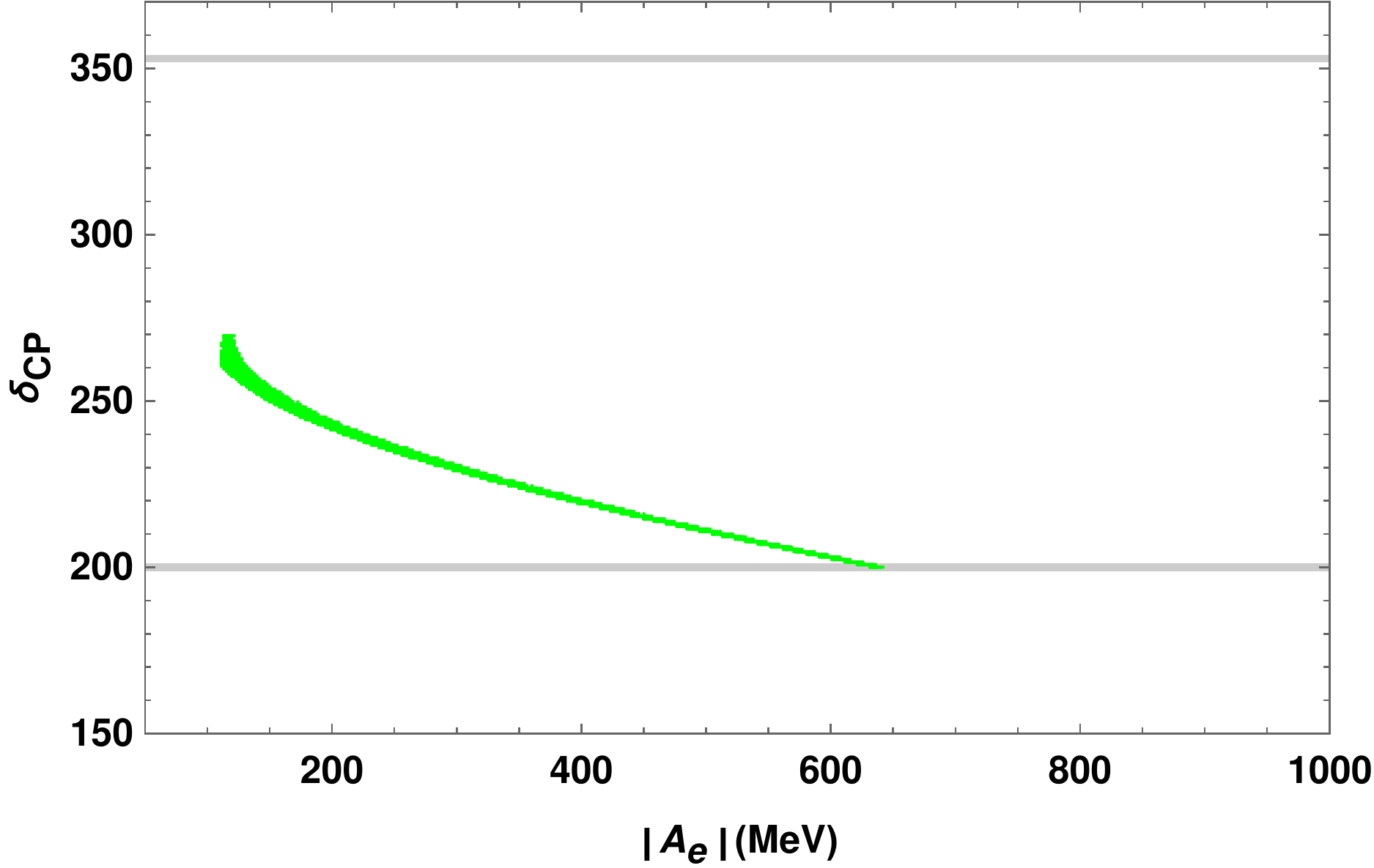}\hspace{5mm}
	\caption{\label{fig:Dvst} 
	Dirac CP-violating phase as function of the $\vert A_{e}\vert$ parameter for the normal and inverted hierarchy, respectively. The solid gray line stands for the experimental data at $3\sigma$.}
\end{figure}

The $\vert A_{e} \vert$ parameter is constrained by the Dirac phase such that its allowed region is reduced substantially in the inverted ordering as we observe in Figure \ref{fig:Dvst}. On the contrary, the region is large for the normal hierarchy and the main reason has to do with the experimental data (see Eqn.(\ref{OND})). Also, we realize a large value for $\vert A_{e} \vert$ is needed to reach the best fit ($194^{\circ}$).

\begin{figure}[ht]
\includegraphics[width=0.4\linewidth]{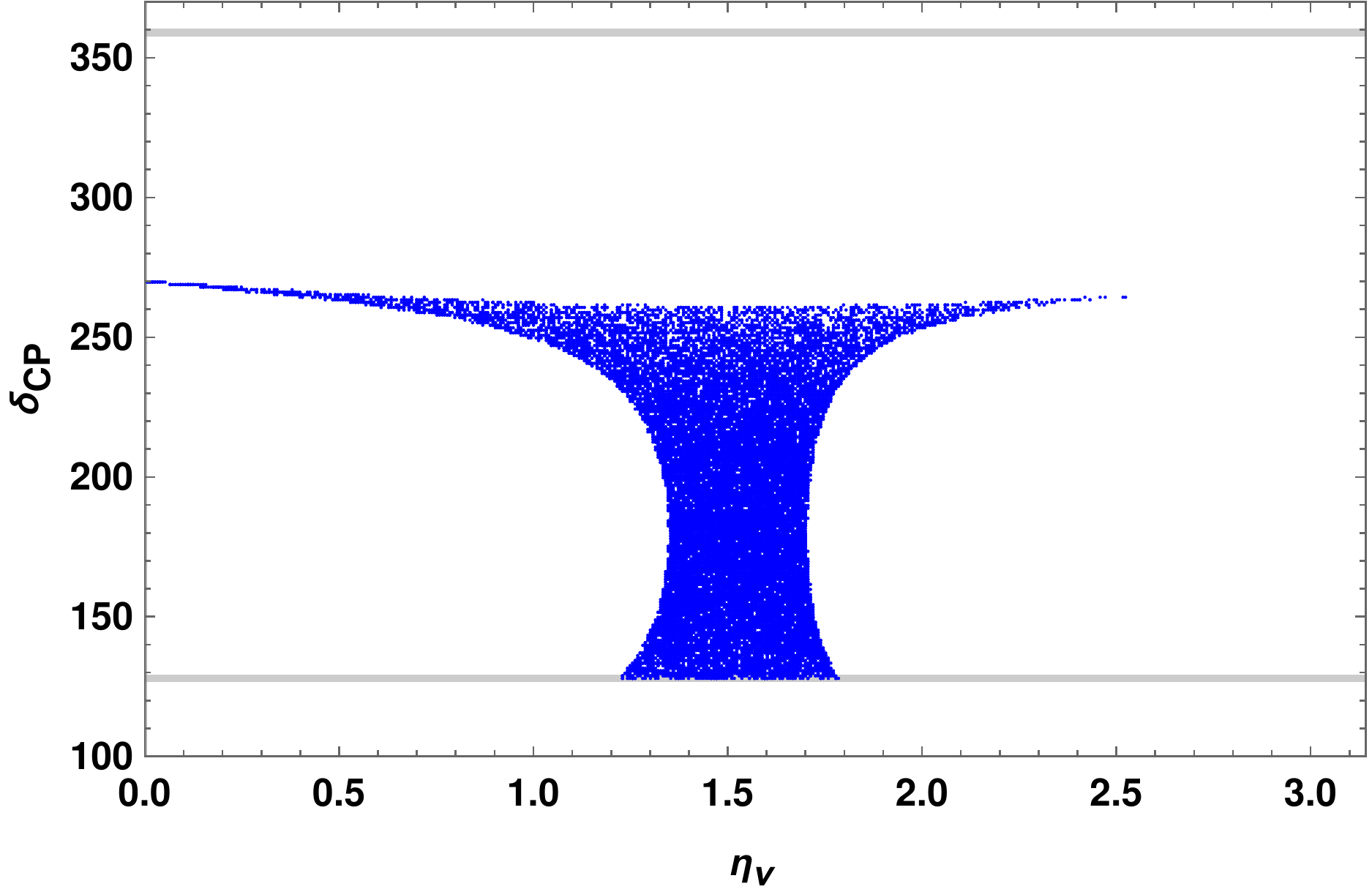}\hspace{5mm}\includegraphics[width=0.4\linewidth]{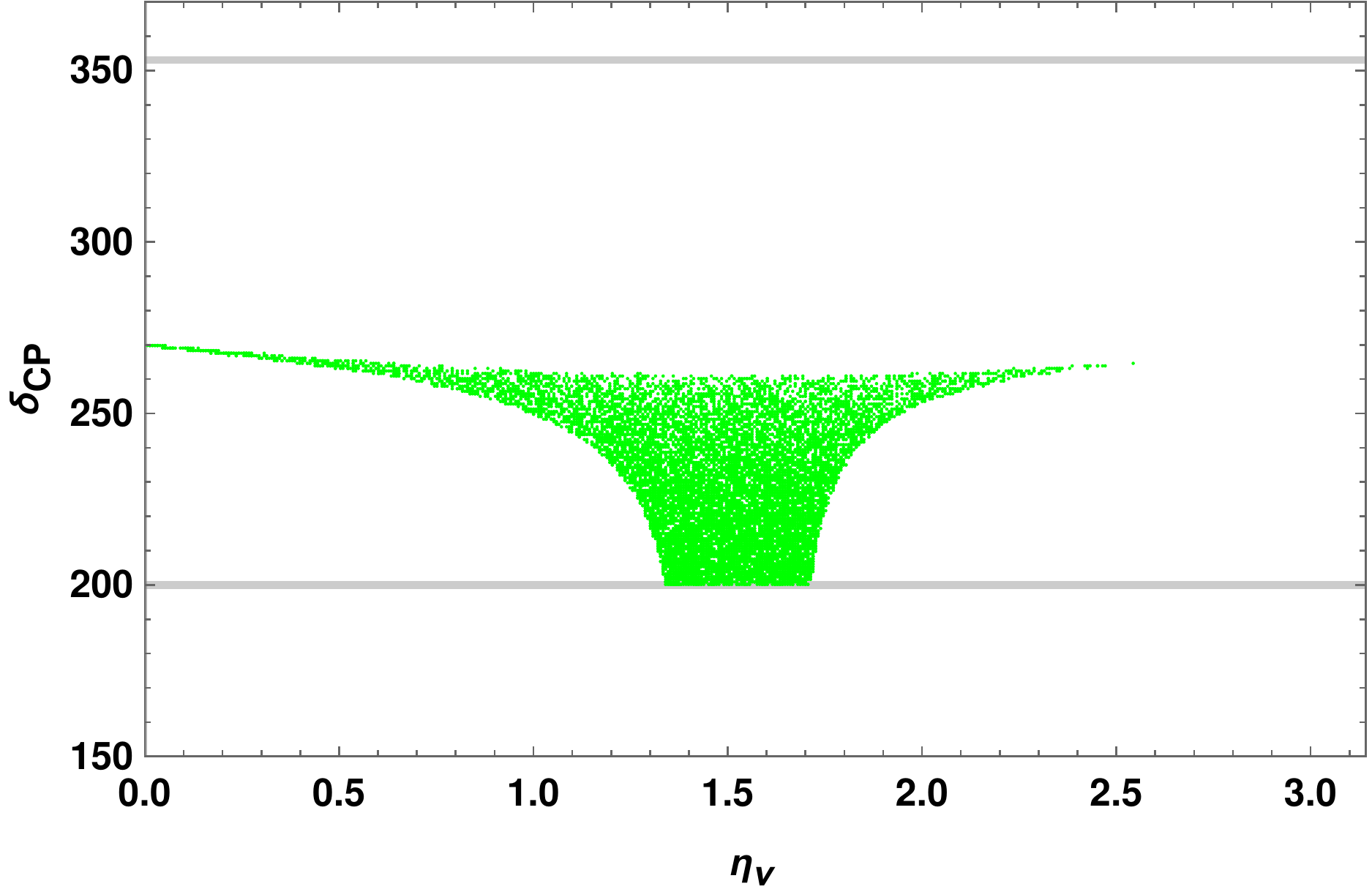}\hspace{5mm}
\caption{\label{fig:DvsE} 
Dirac CP-violating phase as function of the $\eta_{\nu}$ parameter for the normal and inverted hierarchy, respectively. The solid gray line stands for the experimental data at $3\sigma$.}
\end{figure}

In Figure (\ref{fig:DvsE}), the Dirac phase is also show as function of the $\eta_{\nu}$ relative phase. In the inverted ordering the allowed region is smaller than the normal case and the explanation resides on the experimental data at $3\sigma$ for the Dirac phase. 

\section{The neutrinoless double beta decay}
As it is well known, the nature of neutrinos is not well established yet so those can be Dirac or Majorana fermions. This latter can be tested by means of the neutrinoless double beta decay~\cite{Vergados:2002pv,Rodejohann:2011mu,Vergados:2012xy,Pas:2015eia} which occurs if and only if the neutrinos are Majorana fermions. In general, the decay width of $0\nu \beta \beta$ is written as 
\begin{equation}
	\Gamma^{0\nu}=\sum_{j} G_{j}\left(Q,Z\right) \vert \mathcal{M}_{j}F_{j}\vert^{2},
\end{equation}
the above expression includes all the possible mechanisms whose matrix element is denoted by $\mathcal{M}_{j}$ and $F_{j}$ stands for a dimensionless particle physics parameter; $G\left(Q,Z\right)$ is a phase space factor that can also depend on the particle physics~\cite{Vergados:2002pv, Rodejohann:2011mu,Vergados:2012xy, Pas:2015eia}. In the scenario where light active neutrinos are involved the main contribution is provided by $F_{\nu}$ the parameter and decay width are written as
\begin{equation}
\Gamma^{0\nu}=G^{0\nu}_{01}\vert \mathcal{M}^{0\nu}_{\nu}\vert^{2} \vert F_{\nu} \vert^{2}
\end{equation}	
with
\begin{eqnarray}
F_{\nu}=\frac{1}{m_{e}}\overbrace{\sum_{i=1}\left(\mathbf{U}\right)^{2}_{e i} m_{i}}^{m_{ee}}.
\end{eqnarray}
where $\mathbf{U}$ stands for the PMNS matrix, $m_{i}$ being the active neutrino masses and  the $m_{e}$ electron mass. So far, there is an upper limit  $\vert m_{ee}\vert< 0.06-0.2~eV$ given  by GERDA collaboration~\cite{Agostini:2013mzu}.

In our model, $m_{ee}$ is not affected by the contribution of the charged lepton sector. Indeed, this is given by
\begin{equation}
\big|m_{ee}\big|=\bigg| \cos^{2}{\theta_{13}}\left(\vert m_{1}\vert \cos^{2}{\theta_{12}}+\vert m_{2}\vert\sin^{2}{\theta_{12}}\right)+\vert m_{3
}\vert \sin^{2}{\theta_{13}} \bigg|.
\end{equation}

From the neutrino oscillation data~\cite{deSalas:2020pgw,Esteban:2020cvm}, the available information on the neutrino masses is given by the squared mass difference $\Delta m^{2}_{21}= \vert m_{2} \vert^{2}-\vert m_{1} \vert^{2}$ and  $\Delta m^{2}_{31}= \vert m_{3} \vert^{2}-\vert m_{1} \vert^{2}$ ($\Delta m^{2}_{13}= \vert m_{1} \vert^{2}-\vert m_{3} \vert^{2}$) for the normal (inverted)
ordering. At $3\sigma$, we have $\Delta m^{2}_{21}[ 10^{-5} eV^{2}]=6.94-8.14$ and $\Delta m^{2}_{31}[ 10^{-3} eV^{2}]=2.47-2.63$ ($\Delta m^{2}_{13}[ 10^{-3} eV^{2}]=2.37-2.53$). Then, fixing two neutrino masses in terms of the lightest one and the squared mass scale, we get for the normal and inverted ordering
\begin{eqnarray}
\vert m_{3} \vert^{2}&=& \Delta m^{2}_{31}+\vert m_{1} \vert^{2},\qquad \vert m_{2} \vert^{2}= \Delta m^{2}_{21}+\vert m_{1} \vert^{2};\nn\\
\vert m_{2} \vert^{2}&=& \Delta m^{2}_{21}+\Delta m^{2}_{13}+\vert m_{3} \vert^{2},\qquad \vert m_{1} \vert^{2}= \Delta m^{2}_{13}+\vert m_{3} \vert^{2}.
\end{eqnarray}

\begin{figure}[ht]
\includegraphics[width=0.4\linewidth]{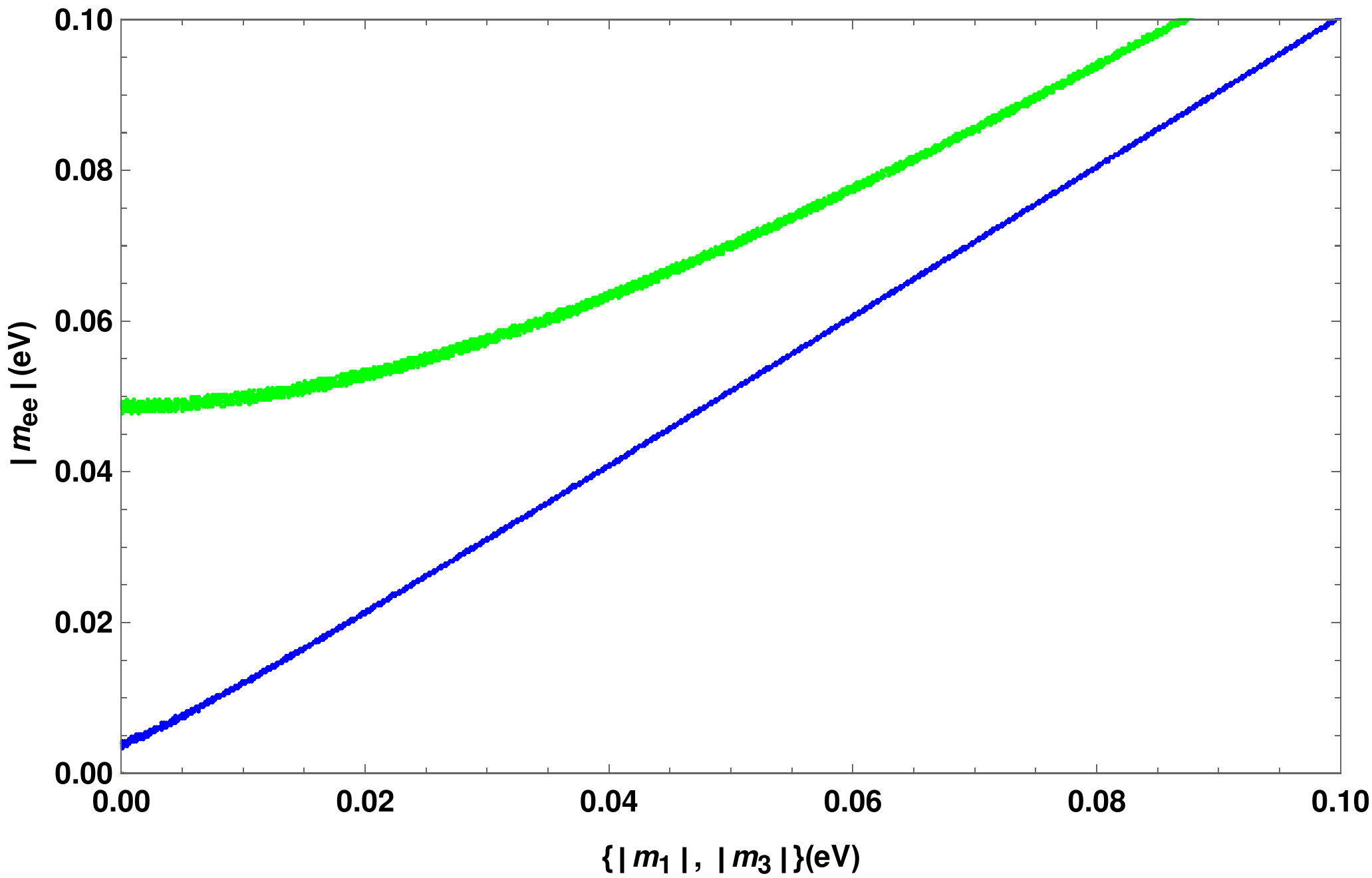}
\caption{\label{fig:mee} 
		Allowed regions for $\vert m_{ee}\vert$ as function of the lightest neutrino mass. }
\end{figure}

Up to now, the absolute neutrino masses are still unknown however there is a bound, $\sum_{i=1} m_{i}<0.12~eV$, given by Planck collaboration that can be useful to determine them. In order to scan the allowed regions of values for the $\vert m_{ee} \vert$ mass, we vary $\theta_{12}$ and $\theta_{13}$ according to Eqn. (\ref{OND}) as well as the $\vert m_{1,3}\vert$ lightest neutrino mass varies up to $0.1~eV$. Our findings are shown in Figure (\ref*{fig:mee}).
As one notices, the allowed regions are compatible with the GERDA phase-I data\cite{Agostini:2013mzu}.

\section{Concluding remarks}
Discrete symmetries have been an option to understand the peculiar lepton pattern. Too much effort has been made in order to build the model, with the ideal flavor symmetry within the suitable framework, that matches the experimental data. Nonetheless, there is no such model so far.

In this letter, we just addressed the lepton masses and mixings by implementing the $\mathbf{S}_{3}\otimes \mathbf{Z}_{2}$ in the $B-L$ gauge model with enlarged scalar sector. A particular choice is realized on the scalar and lepton families under the $\mathbf{S}_{3}$ irreducible representations, and if the Dirac, Majorana and charged lepton Yukawa couplings are real and complex, respectively. This leads to a deviation from Cobimaximal mixing pattern due to the charged leptons. According to our findings, the PMNS mixing matrix does depend on two free parameters ($\vert A_{e}\vert$ and $\eta_{\nu}$) which were constrained by analyzing their effect on the atmospheric angle and the Dirac CP-violating phase that end up being deviated from $\pi/4$ and $3\pi/2$, respectively. For normal and inverted hierarchy, we found a set of values for the free parameters that accommodates the atmospheric angle with great accuracy (up to $3\sigma$) and it can be on the lower and upper octant. We stress that there is not a clear difference between both regions of values for the $\eta_{\nu}$ free parameter. Taking into account
the Dirac CP-violating phase, the numerical study allows to distinguish two regions of values for each free parameter, this can be relevant to test the model in future experiments~\cite{JUNO:2015zny,DUNE:2020ypp,Hyper-Kamiokande:2018ofw}. Additionally, the $\vert m_{ee}\vert$ effective neutrino mass is completely independent of the charged lepton contribution so that it was well fixed by the reactor and solar angles. As a result, the allowed regions are consistent with the available data.

Certainly, this model can be considered like one more to the market with all its limitations. In our opinion, there are still many ideas to explore by utilizing the $\mathbf{S}_{3}$ symmetry in the context of three Higgs doublets where a rich phenomenology has not been explored with detail. For instance, an exhaustive study on the scalar and gauge sector\footnote{A partial study was realized in ~\cite{Gomez-Bock:2021uyu}}  and its effect on the charged lepton flavor violation and leptogenesis.
Consequently, we pretend to be more ambitious so that the quark sector and the above issues will be addressed in the model to have a complete study.

\section*{Acknowledgements}
JCGI thanks Alma Leticia Gálves Mendoza for its valuable support at CECyT No. 16. Special mention to Rosalva Iranel Sánchez García and Edgar Geovanni Trejo Escamilla for preparing the early spanish version  of the manuscript. This work was partially financed by Secretaria de Investigación y Posgrado del Instituto Politécnico Nacional under Projects 20230568 and PAPIIT IN109321.  
\appendix


\bibliographystyle{bib_style_T1}
\bibliography{references.bib}


\end{document}